\documentclass[12pt,preprint]{aastex}

\shorttitle{Thick Disk Origins}
\shortauthors{Brook, Kawata, Gibson, Freeman}

\begin{document}
\include{psfig}
\title{The Emergence of the Thick Disk in a CDM Universe}

\author{Chris B. Brook, Daisuke Kawata, Brad K. Gibson}
\affil{Center for Astrophysics \& Supercomputing, Swinburne
University, Mail \#31, P.O. Box 218, Hawthorn, Victoria, 3122, Australia}
\email{cbrook@astro.swin.edu.au}
\and 
\author{Ken C. Freeman}
\affil{Mount Stromlo Observatory, Australia National University, Western Creek, ACT 2611, Australia}

\begin{abstract}
The disk galaxy simulated using our chemo-dynamical galaxy formation code, {\tt{GCD+}}, is shown to have a thick disk component. This is evidenced by the velocity dispersion versus age relation for solar neighbourhood stars, which clearly shows an abrupt increase in velocity dispersion at lookback time of approximately 8 Gyrs, and is in excellent agreement with observation. These thick disk stars are formed from gas which is accreted to the galaxy during a chaotic period of hierarchical clustering at high redshift. This formation scenario is shown to be consistent with observations of both the Milky Way and extragalactic thick disks.      

\end{abstract}

\section{Introduction}
Since its existence was detected through star counts by Gilmore \& Reid (1983), the thick disk has become established as a component of the Milky Way that is distinct from the thin disk. It is hoped that the  old age of the stars associated with the thick disk of the Milky  Way  means that it will serve as a fossil record of the formation processes of the early parts of Galactic evolution.

Several large surveys have been undertaken to constrain the properties of the Milky Ways thick disk. The local number density of thick disk stars is about 6-13\% of that of the thin disk (e.g. Chen et~al. 2001); the thick disk has a scale height of 0.6-1 kpc (e.g. Phelps et al. 1999, Ojha 2001, Chen et~al. 2001), about three times larger than the thin disk; a scale length of $\sim$ 3 kpc (e.g. Robin~et~al. 1996; Ojha 2001)
compared with a thin disk scale length of $\sim$ 4.5 kpc (e.g. Habing 1988);
its stars are  old, almost exclusively older than 12 Gyrs (e.g. Gilmore \& Wyse 1985; Edvardsson et~al. 1993; Fuhrmann  1998). Thick disk stars have a wide range of metallicity,  $-$2.2$\leq$[Fe/H]$\leq$$-$0.5, (Chiba \& Beers 2000, CB hereafter), although  the metal weak ([Fe/H]$<$$-$1.0) tail of the distribution contributes only $\sim$1\% of the thick disk (Martin \& Morrison 1998), and may be a different population to the canonical thick disk (Beers et~al. 2002). We also note that  Feltzing Bensby \& Lundstr\"{o}m (2003, hereafter FBL)  find that the thick disk metallicity distribution extends to metallicities higher than [Fe/H]=$-$0.5. The metallicity distribution peaks at [Fe/H]$\sim$$-$0.6 (Gilmore \& Wyse 1985; Wyse \& Gilmore 1995; CB). Thick disk stars are also characterised by their warm kinematics, and a rotation that lags the thin disk e.g. Str\"{o}mgren (1987) find ($\sigma_U$,$\sigma_V$,$\sigma_W$)=(65,54,38) km/s, V$_{lag}\sim$40 km/s.

There is mounting evidence that  chemical trends in the thick and thin disk stars  are different (Fuhrmann 1998; Prochaska et~al. 2000; Tautvaisiene et~al. 2001; Schr\"{o}der \& Pagel 2003; Mashokina et~al. 2003, MGTB hereafter; Reddy et~al. 2003; FBL; but see also  Chen et~al. 2000 for a conflicting view). A major diagnostic coming from such different chemical trends between thin disk and thick disks is the different $\alpha$-element to iron abundance ratios, indicating different formation timescales. The enhanced $\alpha$ element abundances compared to iron which are observed in thick disk stars indicate a short formation timescale in which enrichment is dominated by Type II Supernovae (SNe II). Although FBL found evidence of SNe~Ia enrichment in the most metal rich of the thick disk stars, their values for [$\alpha$/Fe] are systematically higher than thin disk stars with the same [Fe/H]. Fuhrmann (1998), Gratton et~al. (2000), and Mashonkina \& Gehren (2001) conclude that star formation in the thick disk lasted less than 1 Gyr, while correlations of various chemical elements in thick disk stars lead MGTB to estimate this time scale to be $\sim$0.5 Gyrs. MGTB also conclude that the homogeneity of abundances of thick disk stars indicates that it formed from gas which was well mixed,  a conclusion which is consistent with  the results of Nissen \& Schuster (1997), Fuhrmann (1998), and Gratton et~al. (2000). 

Over the past decade or more, N-body cosmological simulations have been successful in reproducing many properties of disk galaxies. Recently, using a new implementation of SNe feedback, Brook et~al. (2003, BKGF hereafter) successfully simulated a  realistic disk galaxy which has its stellar mass dominated by a young stellar disk component, surrounded by a less massive, old, metal poor, stellar halo. In this letter we examine this simulated disk galaxy for evidence of a thick disk component, looking for clues as  to the origins of this component within the hierarchical galaxy formation of a CDM context.  

     In Section~2, we give details of our N-body chemo-dynamical evolution code, {\tt{GCD+}}, and the semi-cosmological galaxy formation model that we employ. Initial conditions are chosen that  lead to the formation of a late type galaxy, whose properties are presented in Section~3. Further study of our simulated galaxy provides evidence for the existence of a distinct thick disk population, similar to that of the Milky Way. This allows us, in Section~4, to trace the major thick disk formation epoch and propose a thick disk formation scenario. In Section~5 we compare our scenario to other theories of thick disk formation, and discuss how current observational data may be used to distinguish these theories.    

\section{The Code and Model}
This paper analyses data from the simulated late type galaxy of the {\it{Adiabatic Feedback Model}} from BKGF. We briefly review some features of the code and model. 
Our galactic chemo-dynamical evolution code, {\tt GCD+}, self-consistently models
the effects of gravity, gas dynamics, radiative cooling, and star
formation. We include SNe~Ia and SNe~II feedback, and trace the lifetimes of individual stars  which enables us to  monitor the 
chemical enrichment history of our simulated galaxies. 
Details of {\tt GCD+} can be found in Kawata \& Gibson
(2003).

We assume that 10$^{51}$ ergs is fed back as thermal energy from each SNe. The energy is smoothed over the nearest neighbour gas particles using the SPH kernel. This feedback scheme is known to be inefficient (Katz 1992). To address this problem, gas within the SPH smoothing kernel of SNe~II explosions is prevented from cooling, creating an adiabatic phase for gas heated by such SNe. This is similar to a model presented in Thacker \& Couchman (2000)

The semi-cosmological version of {\tt GCD+} used here is based upon the
galaxy formation model of Katz \& Gunn (1991).  The initial condition is
an isolated sphere of dark matter and gas. This top-hat overdensity has an amplitude, $\delta_i$, at initial redshift, $z_i$, which is approximately related to the collapse redshift, $z_c$, by $z_c$=0.36$\delta_i$(1+$z_i$)$-$1 (e.g. Padmanabhan 1993). We set $z_c$=1.8, which determines $\delta_i$ at $z_i$=40.         Small scale density
fluctuations are superimposed on the sphere, parameterized by $\sigma_8$.  These perturbations are the seeds for local collapse and subsequent star
formation. Solid-body rotation corresponding to a spin parameter,
$\lambda$, is imparted to the initial sphere to incorporate the effects of longer wavelength fluctuations. A large value of $\lambda$ is chosen along with initial conditions in which no major merger occurs in late epochs ($z<$1), in order to ensure a disk galaxy is formed. For the flat CDM model described
here, the relevant parameters include  $\Omega_{0}$=1, baryon fraction, $\Omega_{b}=0.1$, $H_0=50$km s$^{-1}$Mpc$^{-1}$, total mass,  $5\times
10^{11}$~M$_\odot$, star formation efficiency, c$_*$=0.05, spin
parameter, $\lambda$=0.0675 and $\sigma_8$=0.5. 
We employed 38911 dark matter and 38911
gas/star particles, making the resolution of this study comparable to 
other recent 
studies of disk galaxy formation (e.g. Abadi et~al. 2003).

\section{Properties of Final Galaxies}

Fig.~1 shows a density plot of the final stellar and gas  populations of the simulated galaxy, both face on and edge on. The stellar mass is dominated by  a young disk component. A large gaseous thin disk, still undergoing star formation, has also formed. The stellar population of the low mass stellar halo component is  old and metal poor.   The star formation history is shown in Fig.~2, which plots look back time against the star formation rate (SFR). The SFR peaks $\sim$9 Gyrs ago. The SFR then declines over the next 4-5 Gyrs, but is reasonably steady over the last $\sim$5 Gyrs.

We are interested here in the thick disk component. In observations of the 
Milky Way, the peculiar velocity of stars is a useful diagnostic to 
distinguish thick disk stars from thin disk stars. For observed Solar 
neighbourhood stars, it is well established that there is a relationship 
between velocity dispersion and age. Fig.~3 shows the observed relation as 
grey symbols, with error bars, as read from Fig.~3 of Quillen \& Garnett 
(2001), who use the data of Edvardsson et~al. (1993). The velocity dispersions
plateau between the ages of $\sim$10 and $2$ Gyrs; these stars belong to the 
old thin disk. A slight decrease in dispersion is apparent for young star 
particles,  less than $\sim$2 Gyrs old. Notably, the  velocity dispersions 
increase abruptly, approximately doubling,  $\sim$10 Gyrs ago. These older 
stars belong to the thick disk{\footnote{We note that a more recent study by 
Nordstrom et~al. (2004) revisits this relation, and finds that the thin disk 
stars do show a slight increase of velocity dispersion with age, but stars 
older than $\sim$10 Gyrs were excluded from their study.}. 
In Fig.~3 we use ``$+$'' symbols to plot the relation between the three 
components of the velocity dispersion, and the stellar age, for solar 
neighbourhood star particles in our simulation. The  solar neighbourhood is 
defined as the annulus 4$<R_{XY}<$12 kpc and $|Z|<$1 kpc.   
The abrupt increase in velocity dispersion, associated with the thick disk,
is unmistakably reproduced in our simulation.

Interestingly, the timing of the  abrupt increase in velocity dispersion in our simulations corresponds to the end of the peak of the SFR (Fig.~2). This motivated us to explore the main processes which were taking place during this epoch of the peak star formation rate, when stars of age $>$8 Gyrs  were forming rapidly. This epoch appears to be associated with thick disk formation. Fig.~4 displays  the evolution of the stellar (upper panels) and gaseous (lower panels) density distribution  of the simulated galaxy during this epoch, between 9.5 and 8.3 Gyrs ago. This epoch is characterised by a series of merger events, and is by far the most chaotic period during the galaxy's evolution. At the beginning of this epoch, $\sim$10 Gyrs ago, at least four proto-galaxies of significant mass exist. These building blocks are gas rich, with combined gas mass of $\sim$2.4$\times10^{10}$M$_\odot$, compared with stellar mass of $\sim$7.0$\times10^{9}$M$_\odot$. By $\sim$8 Gyrs ago, a single central galaxy has emerged. The flattened nature of this central galaxy is  apparent by the last panel of Fig.~4, yet this disk is significantly thick. The thin disk forms in the quiescent period of the remaining 8 Gyrs.

There is an offset in the timescales  of the thick disk formation between the simulation and observation, with the observational (simulated galaxy's) abrupt increase in velocity dispersion occurring around 10 (8) Gyrs ago. 
To further investigate this offset, we examine a simulation which evovles to a disk galaxy in which the collapse redshift, $z_c$=2.2, was earlier than for the model examined so far in this paper. This galaxy less resembles the Milky Way in that it has a large bulge component, yet it also has a prominant disk. The major merging epoch of this galaxy is earlier than for our previous model, and this is reflected  in the relation between velocity dispersion and age represeted by ``$\times$'' in Fig.~3, where the abrupt increase in velocity dispersion has been pushed back in time to $\sim$10 Gyrs. The implications is that the  
offset is a result of the merging histories of the different galaxies.

We consider some further properties of the stars which are formed in the 
chaotic period of galaxy formation which we have identified with
thick disk formation, and make comparisons with stars which 
form in the later quiescent epoch. We  remind the reader that our simulated 
galaxy is not a model of the
Milky Way, but rather a Milky-Way like late type galaxy: 
precise quantitive agreement in any of the properties examined would 
not be expected.  
 Star particles with
age $<$7 (8.5$<$ age $<$10.5) Gyrs are referred to as thin (thick) disk stars. 
A (somewhat arbitrary) cut in rotational velocity, ($V_{rot}$$<$50 km/s) is 
then made, and all stars with  ($V_{rot}$$>$50 km/s) are defined as halo stars.
This cut does not affect the thin disk, 
but identifies around $10\%$ of stars which formed during the epoch of thick 
disk star formation as halo stars.
 The simplicity of our assignment
of stars to thin and thick disks on the basis of age, allows us to 
link properties of the different components stars to processes 
occuring at the different epochs at which they form.
 
Fig.~\ref{fig:scalel} shows the surface density distributions of the thin and
thick disk components of our simulated galaxy. 
Both components are well approximated by exponential
profiles within the region 4$<$R$<$12 kpc,
with scale lengths of 4.1 kpc and 2.6 kpc for the thin and thick disks 
respectively.  Fig.~\ref{fig:scaleh} shows the vertical density 
distributions of the thin and thick disk components, averaged
over the region 4$<$R$<$12 kpc. The exponential profile of these plots 
indicate scale heights of 0.5 kpc and 1.3 kpc for the thin and thick disk
respectively.  The rotation 
curves of the thin (thick) component is
 shown  as a dashed (dot-dashed) line in Fig.~\ref{fig:rot}.
The thick disk lags the thin disk by approximately 20 km/s in
the solar neighbourhood. The implication of these plots is to
confirm that our simulated galaxy has both a thin and thick disk component 
and that these components are qualitatively similar to
those of the Milky Way (see Section~1).

The metallicity distribution functions of solar 
neighbourhood stars, defined as above,
are shown in Fig.~\ref{fig:mdf}.
Thin disk stars (dashed line) peak at [Fe/H]$\sim -$0.1.
Thick disk stars (dot-dashed line) have greater spread in
metallicity than thick disk stars, despite their tighter age spread, 
and peak at [Fe/H]$\sim -$0.3.
This makes our thick disk component more metal rich than that
of the Milky Way. In Brook et~al. (2004)
a link was made between the mass ratios of the components of simulated 
galaxies, to the metallicties of those components.  
The stellar halo of our simulated galaxy is
both more massive and more metal rich than that observed 
for the Milky Way (Brook et~al. 2004). This would naturally
be associated with the gas rich building blocks associated with
thick disk formation  in 
our simulation being more metal enriched than those of the Milky Way.
Halo stars (dotted line) are defined as stars with $V_{rot}<50$km/s
and peak at [Fe/H]$\sim -$0.8\footnote{The metallicity
distribution functions derived using the same simulated galaxy in Brook
et~al. (2004) use different methods to define halo and disk stars,
resulting in slightly different MDFs}, with a large tail toward lower
metallicities. As noted, the higher metallicity of the halo stars compared to
the Milky Way, which has a peak at  [Fe/H]$\sim -$1.5, is due 
to the stellar halo of the simulated galaxy being more massive than
the Milky Ways halo. This massive
stellar halo, resulting in larger than observed 
contamination of the solar neighbourood by 
halo stars, may also explain why the
abrupt increase in velocity dispersion seen in Fig.~3 is larger
for our simulated galaxy than for the Milky Way.

\section{Discussion}
We have shown that properties of solar neighbourhood
star particles from our simulated 
disk galaxy are intimately related to processes which can be
associated with different epochs. The majority of solar neighbourhood
stars which form during a 
chaotic merging period, between 10.5 and 8.5 Gyrs ago in our simulation, 
are characteristically thick disk stars. A small portion of stars
born in this time are halo stars.
Solar neighbourhood stars which form in the 
more quiescent period over the past $\sim$8 Grys  are 
characteristically thin disk stars.

Our results indicate that the thick disk is created in an epoch of multiple mergers of gas rich building blocks, during which a central galaxy is formed. The stars which form during this merging period, when a high SFR is triggered (Fig.~2), are the dominant source of thick disk stars.  A significant fraction of the gas accreted to the central galaxy during this epoch carries the angular momentum of the proto-galactic cloud, imparted by the large scale structure of the Universe. This angular momentum results in the rotation and flattening of the central galaxy which is formed. Yet this epoch is characterised by violent interactions, and the forming gas disk can be described as dynamically hot, resulting in the high velocity dispersion of the forming stars in the disk. The result is the formation of a thick disk. In falling gas after this multiple merger epoch settles smoothly to a thin disk.     

Several formation scenarios for the thick disk were presented by Gilmore 
et~al. (1989): 
1) a slow, pressure supported collapse which follows the 
formation of halo Population~II stars (Larson 1976); 
2) Violent dynamical heating of the early thin disk by satellite accretion 
(Quinn et~al. 1993) or 
violent relaxation of the Galactic potential (Jones \& Wyse 1983); 
3) accretion of thick disk material directly (Statler 1988; Bekki \& 
Chiba 2000; Abadi et~al. 2003);  
4) enhanced kinematic diffusion of the thin disk stellar orbits (Norris 1987); 
5) a rapid dissipational collapse triggered by high metallicity 
(Wyse \& Gilmore 1988). 
To unravel which of these processes was the major driver of thick disk 
formation, information of the metallicity, ages, and chemical abundances of 
thick disk stars can be compared to the predictions that the various scenarios 
make. 

Lack of a vertical metallicity gradient in the thick disk (Gilmore et~al. 1995), as well as a lack of  large intermediate age thick disk population seems to rule out the slow collapse of scenario 1). The lack of overlapping age distributions between thick and thin disk stars, as well as the discontinuity in their chemical abundances argues, against the enhanced kinematic diffusion scenario 4). Further evidence against such a scenario is the size of the  discontinuity in velocity dispersion, which appears to be  too great to be explained by the known heating mechanisms, i.e. local gravitational perturbations in the thin disk, such as giant Molecular Clouds (Spitzer \& Schartzschild 1953), or transient spiral structure (Barbanis \& Woltjer 1967; Carlberg \& Sellwood 1985). Freeman (1991) suggested that these disk heating  mechanisms saturated or became inefficient at $\sigma\sim$30 km/s. Lack of correlation between the scale length of the thick and thin disk components of disk galaxies observed  in Dalcanton \& Burnstein (2002, DB hereafter), also contradicts scenario 4) for thick disk formation in other galaxies. 
If the thick disk were made up of primarily of material accreted slowly over time from many smaller satellites, on suitable orbits, as in scenario 3), then the metallicity of the stars would be too low to explain the observed peak metallicity of the thick disk. The well known metallicity-mass relation of galaxies would require accretion of massive satellites ($>$10$^{10}$M$_\odot$), which would destroy the thin disk, in order to obtain metallicities of [Fe/H]$\sim$$-$0.6, unless the accreted satellites are gas rich and the accretion process induces a significant self-enrichment (Bekki \& Chiba 2002). In addition, recent high resolution spectroscopic observations of individual stars in dwarf spheroidals (dSphs), find ~solar $\alpha$ element to iron abundance ratios (e.g. Shetrone et~al. 2001, 2003), which differs significantly from observed [$\alpha$/Fe] and [Fe/H] in the thick disk. This indicates that the accretion of systems similar to dSphs is not the dominant source of thick disk stars.    
As noted by Wyse (2004), all the thick disk formation scenarios 
allow the possibility of a portion of the thick disk stars to have
originated  directly from accretion.

The heating of the disk early in its violent evolution, as in scenario 2), is well supported by Galactic observations (Quillen \& Garret 2001; Wyse 2000; Gilmore et~al. 2002; Freeman \& Bland-Hawthorn 2002; FBL). This scenario is consistent with the observed abrupt increase in velocity dispersions, the distinct chemical properties of the two disk components, and the homogeneity of thick disk abundances. 
We contend that our scenario of thick disk formation during the epoch of multiple mergers of gas rich building blocks is consistent with scenario 2), and hence with these observations of the Milky Ways thick disk.
We note that the scenario 2) suggests two possible heating mechanisms. In the first mechanism of scenario 2), a thin disk forms and is heated by an accretion event. Evidence that the thick disk has had a more intense star formation history than the thin disk (e.g. Bensby et~al. 2003; MGTB) would seem to favour the mechanism we propose over this ``puffed up'' thin disk, although  the latter 
mechanism  is not inconsistent with a rapid thick disk formation timescale (e.g. FBL). Our simulations more closely resembles the mechanism of Jones \& Wyse (1983), in which a thick disk forms during a violent relaxation of the galactic potential, prior to the formation of the thin disk.  We note also that the in situ formation of the thick disk in our scenario leads to efficient self-enrichment. Our scenario also incorporates features of scenario 3), but rather than accreting stars into the thick disk directly, our scenario envisages 
star formation being triggered by accretion of gas rich building blocks
 during a chaotic merging epoch at {\it{high redshift}}.
 The increased cooling due to metals, as mentioned in scenario 5), helps ensure a rapid star formation rate in the thick disk. Thus, although we are proposing
our thick disk formation scenario as ``new'', it 
shares features with previously proposed scenarios.

 We have underway a program to examine
a statistically significant number of simulated disk galaxies similar
to those examined here, with varying masses, angular
momentum and merging histories.
This will help determine whether the
properties of our thick disk-like component 
depend on the adopted model parameters and specific merger histories.
We hope to
predict the expected frequency of 
thick disks and to characterise thick disk properties such as age, metallicity 
and colour of the stellar populations. 

  The formation of the thick disk in the manner we have outlined is a natural consequence of the early, violent merging epoch of the CDM universe, indicating that thick disks would be a widespread component of disk galaxies.   
 Recent, deep observations of a sample of 47 nearby edge-on galaxies by DB indicate that thick disks are almost ubiquitous around disk galaxies. A lack of correlation between thick and thin disk scale lengths, as found by DB,  is also explained by our formation scenario.
Confirmation of these finding would strengthen the case for our  thick disk formation mechanism.

\acknowledgments 
This study has made use of the
Victorian and Australian Partnerships for Advanced Computing, the latter
through its Merit Allocation Scheme.  This work is supported financially
by the Australian Research Council. CBB is funded by an Australian 
Postgraduate Award. We thank Alice Quillen, Mike Beasley and Tim Connors 
for helpful suggestions

\clearpage

\begin{figure*}
\plotone{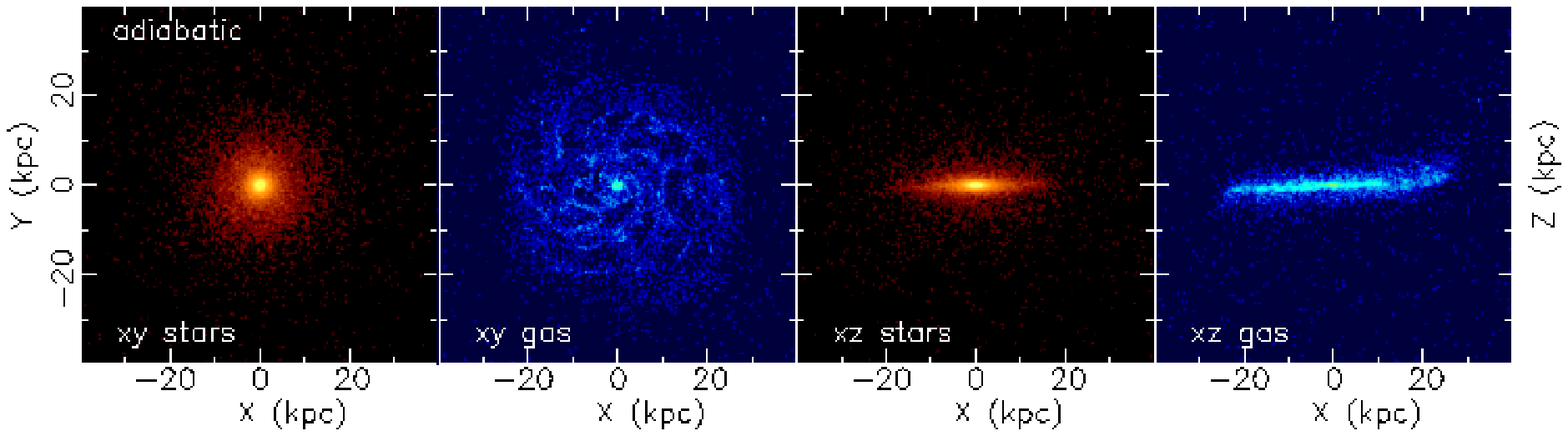}
\caption{ $z$=0 density plots for stars and gas in both the  XY (face on) and XZ (edge on) planes.  The galaxy is dominated by a stellar disk and a large gaseous thin disk, still undergoing star formation, has also formed.   \label{fig1}}
\end{figure*}

\clearpage
\begin{figure}
\plotone{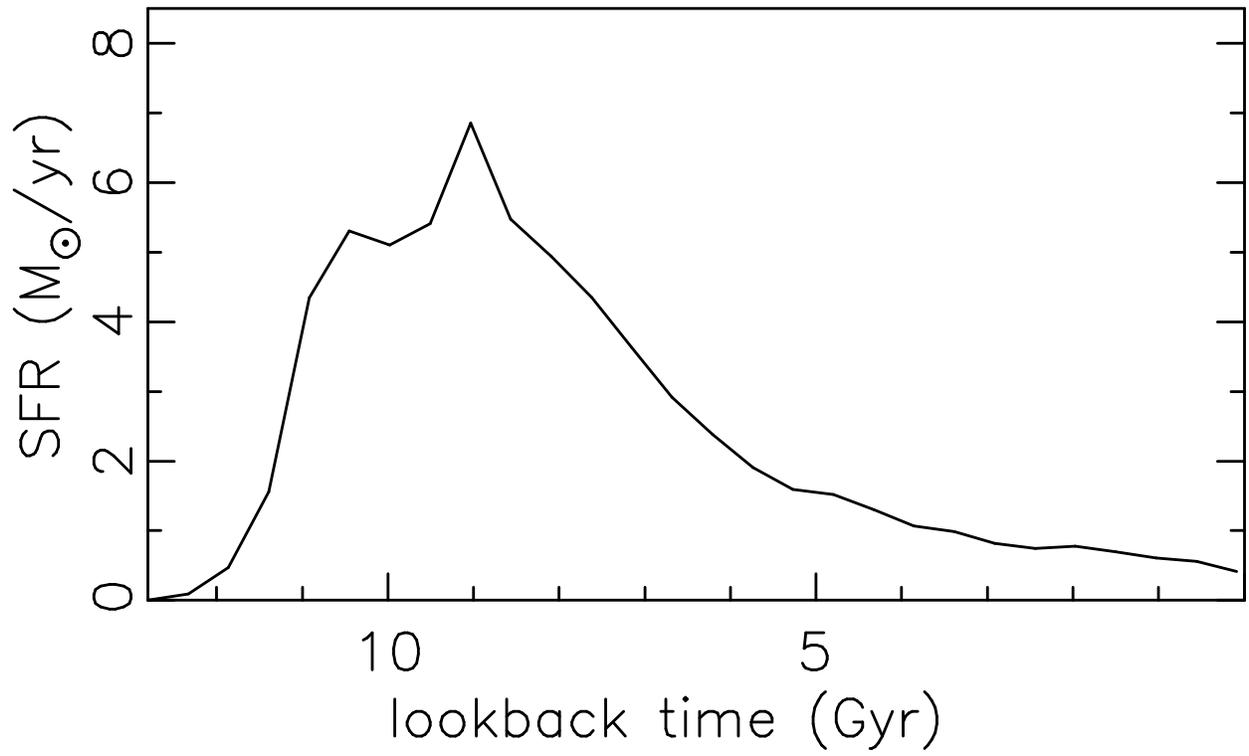}
\caption{Global star formation rate (SFR) as a function of lookback time for our simulation. A peak of star formation is evident $\sim$9 Gyrs ago.
 }
\end{figure}

\clearpage
\begin{figure}
\plotone{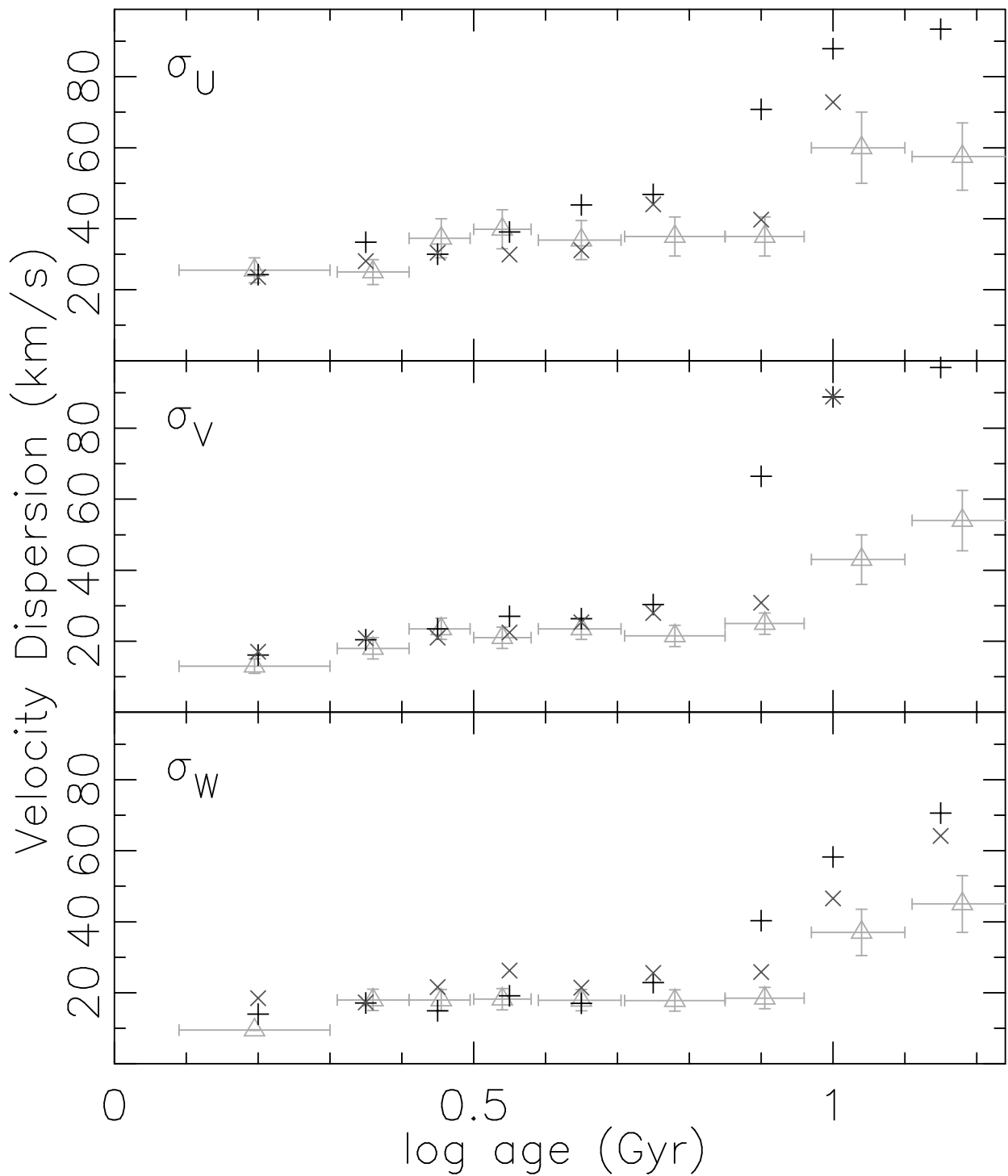}
\caption{Age-velocity dispersion relation in the U, V and W direction for Solar neighbourhood simulation star particles, plotted as ``$+$''. An abrupt increase in the velocity dipersion is apparent at lookback time of $\sim$8 Gyrs. Also plotted, by triangles with error bars, is the velocity dispersion-age relation derived by Quillen \& Garnett (2001, Fig.~3), for the observed solar neighbourhood stars from the sample of Edvardsson et~al. (1993). Results of a disk galaxy simulation with $z_c$=2.2 are plotted as ''$\times$''. In this simulation, the major merging epoch was earlier, and the abrupt increase in velocity dispersion pushed back in time to $\sim$10 Gyrs.}
\end{figure}

\clearpage
\begin{figure*}
\plotone{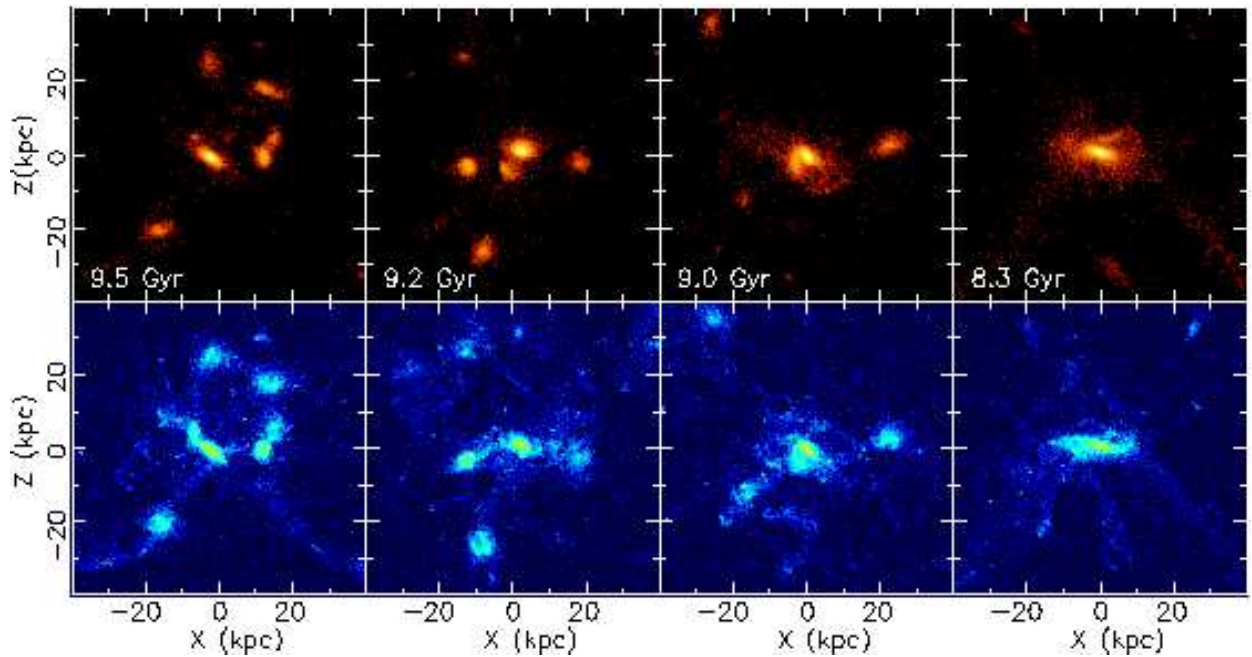}
\caption{Density plot of the evolution of the galaxy's stars (top panel) and gas (lower panels) during the era in which the break in the velocity dispersion plot indicates that thick disk stars are formed. This epoch is charaterised by multiple mergers of gas rich building blocks, resulting in the formation of a central galaxy. } 
\end{figure*}


\clearpage
\begin{figure*}
\plotone{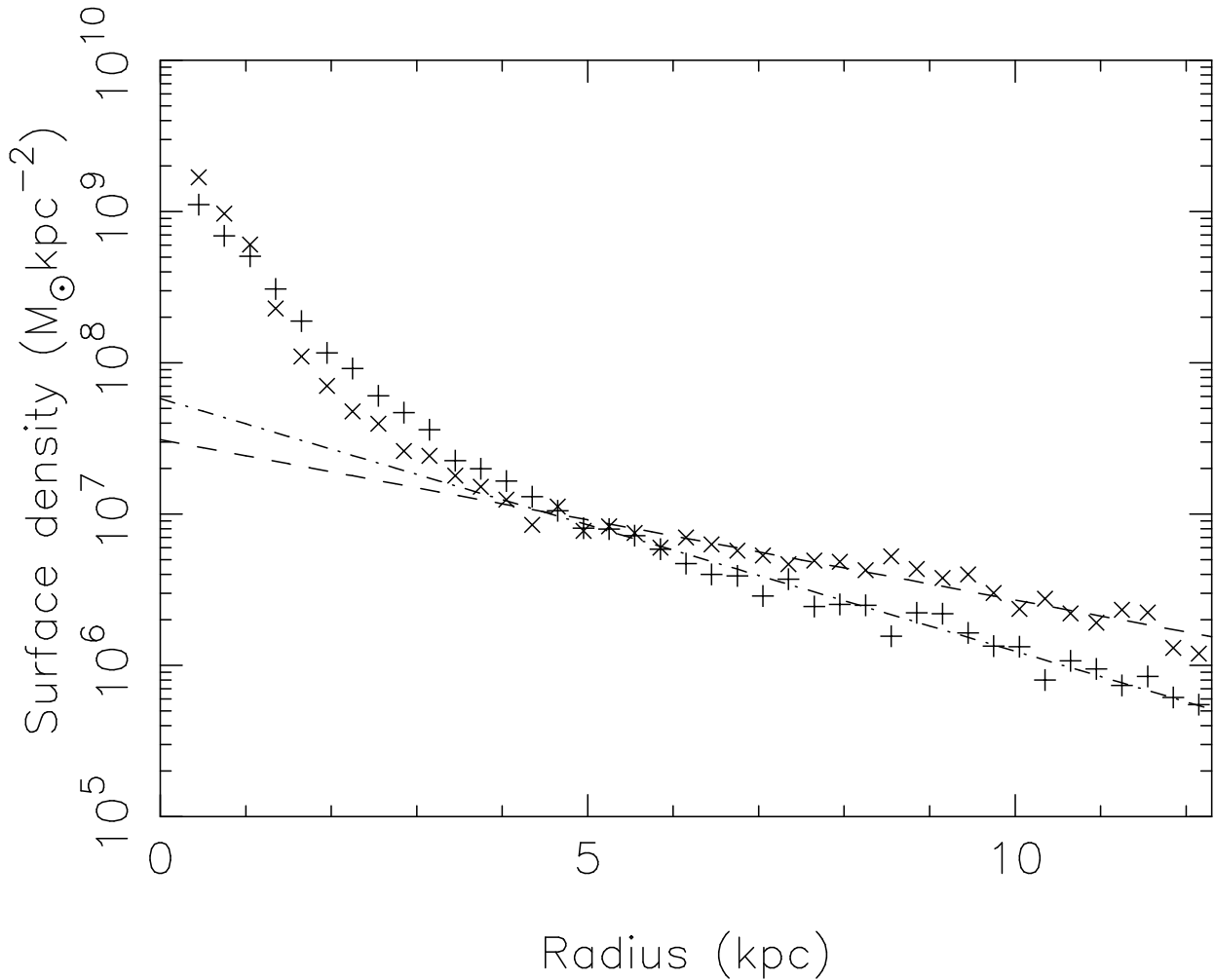}
\caption{Surface density profile of the thin and thick disks, shown as 
``$\times$''s and ``$+$''s respectively. Star particles with ages
$<$7 Grys are called thin disk stars, while star particles with ages 
8.5$<$age$<$10.5 are thick disk stars, with a velocity cut taken at 
$V_{rot}$$>$50 kpc which eliminates halo star contamination. 
The profiles are well approximated
by exponentials,  with fits between 4$<$R$<$12 kpc shown as dashed (dot-dashed)
for thin (thick) disk components indicating scale lengths of
4.1 (2.6) kpc. 
 } 
\label{fig:scalel}
\end{figure*}
\clearpage
\begin{figure*}
\plotone{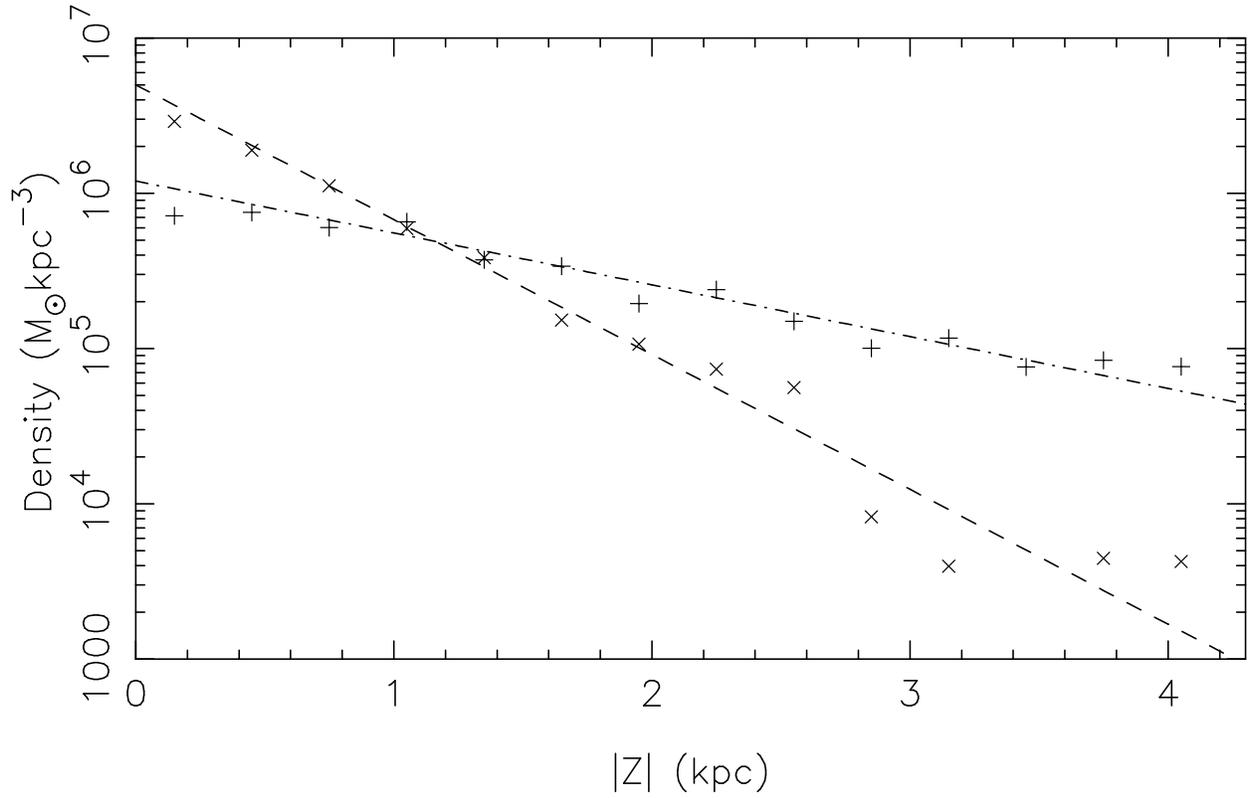}
\caption{Vertical density profile of the thin and thick disks, averaged over 
the region 4$<$R$<$12 kpc, shown as ``$\times$''s and ``$+$''s respectively.
 The profiles are well approximated by exponentials,  with fits
shown as dashed (dot-dashed) lines
for thin (thick) disk components indicating scale heights of
0.5 (1.3) kpc. 
 } 
\label{fig:scaleh}
\end{figure*}
\clearpage
\begin{figure*}
\plotone{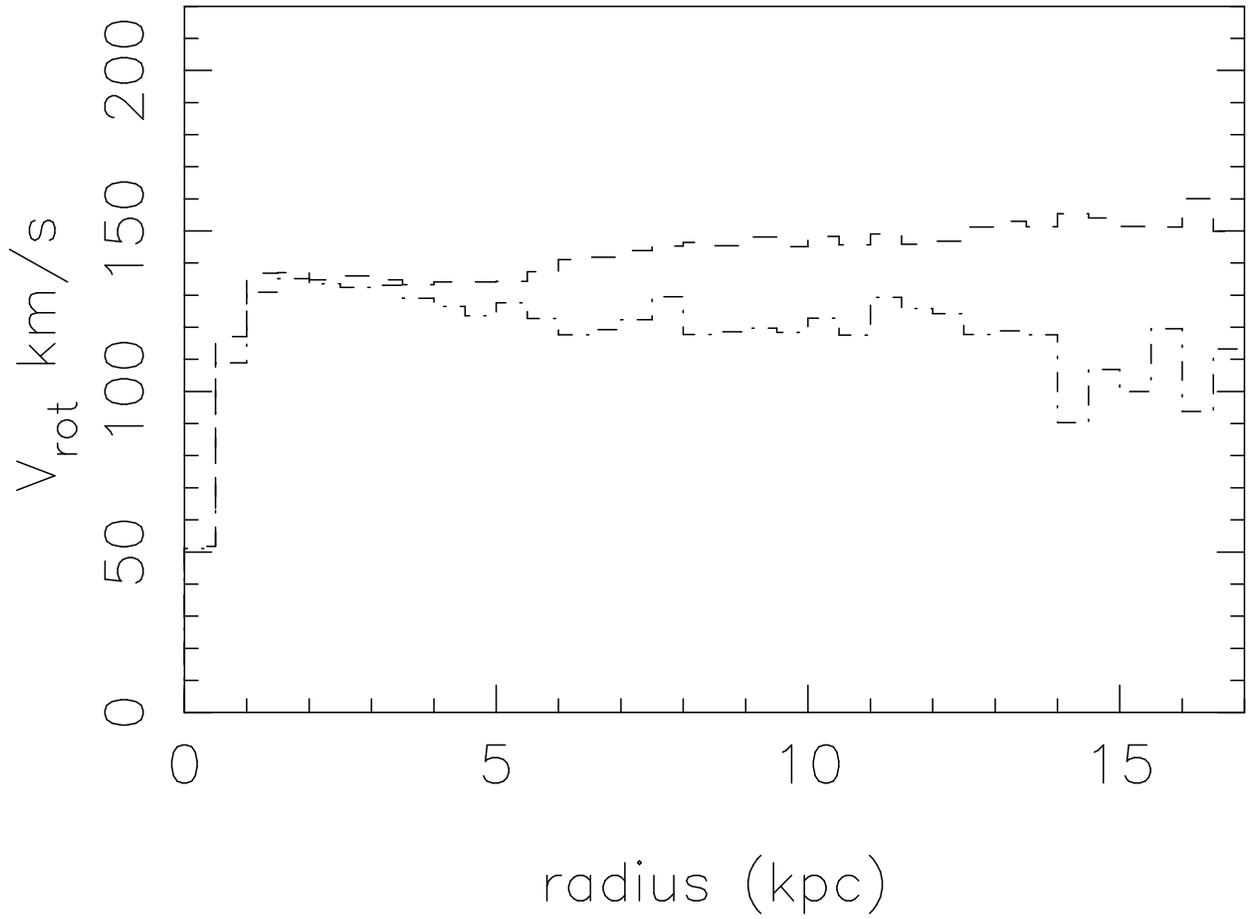}
\caption{Rotation curve for the thin (dashed) and thick (dot-dashed) disk
components.
The thick disk lags the thin disk by approximately 20 km/s in
the solar neighbourhood.
 } 
\label{fig:rot}
\end{figure*}
\clearpage
\begin{figure*}
\plotone{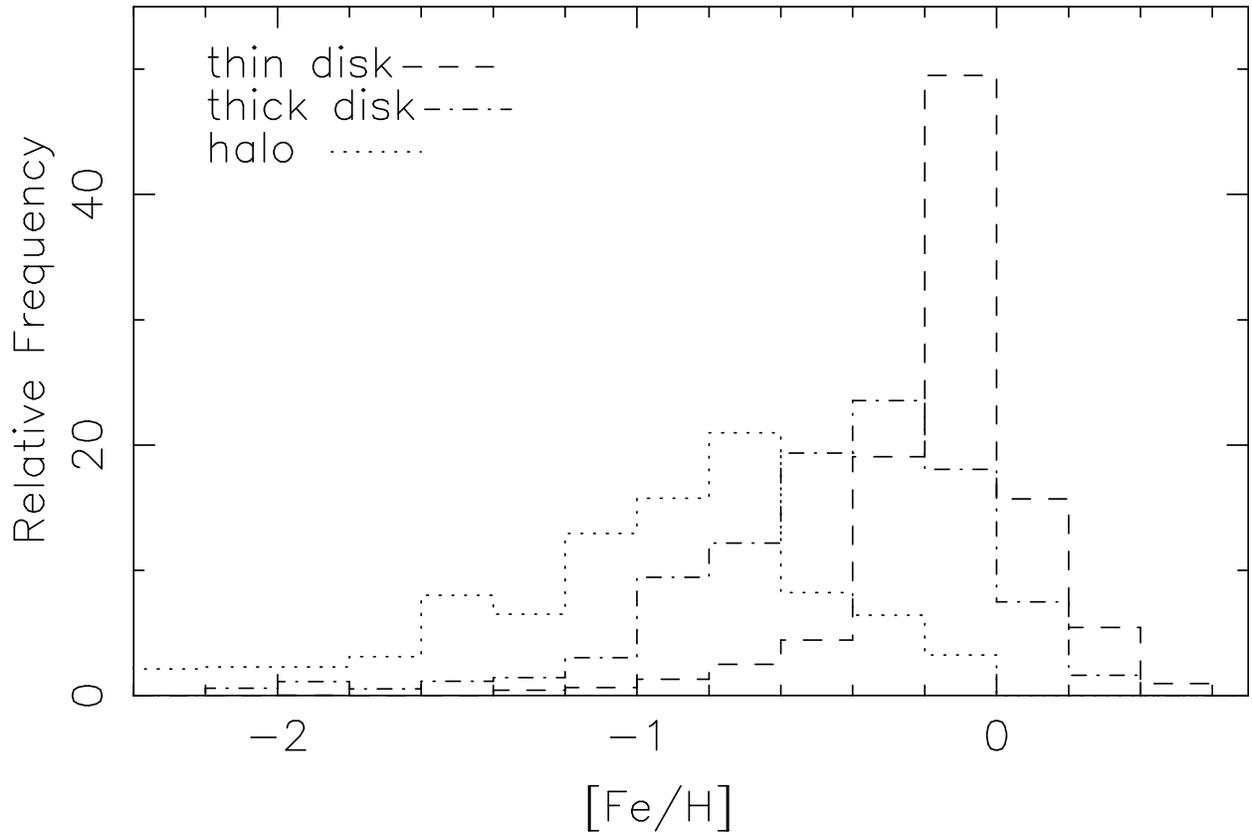}
\caption{The metallicity distribution function of solar 
neighbourhood stars, defined as 4$<$R$<$12 kpc and $|Z|<$1 kpc,
for thin disk stars (dashed line), thick disk stars (dot-dashed),
and halo stars (defined as stars with $V_{rot}<$50 km/s with no age cut, 
dotted line).} 
\label{fig:mdf}
\end{figure*}


\begin{references}
Abadi, M.~G., Navarro, J.~F. \& Steinmetz  M. 2003, ApJ, 597, 21

Barbanis, B.; Woltjer, L.  1967 ApJ, 150 461


Beers, T.~C., Drilling, J.~S., Rossi, S., Chiba, M., Rhee, J., F\"{u}hrmeister, Birgit, Norris, J.~E. \& von Hippel, T. 2002 AJ 124 931

Bekki, K. \& Chiba, M. 2000, ApJ, 534, 89

Bekki, K. \& Chiba, M. 2002, ApJ, 566, 245

Bensby, T., Feltzing, S.,  \& Lundstr\"{o}m, I. 2003, A\&A, 410, 527

Brook, C.~B,  Kawata, D., Gibson, Brad~K., \& Flynn~C. 2004, MNRAS, {\it{in print}}

Carlberg, R. G. \& Sellwood, J. A. 1985 ApJ, 292, 79

Chen, Y.~Q., Nissen, P.~E., Zhoa, G., Zhang, H.~W., \& Benoni, T. 2000, A\&AS, 141, 491

Chen, B., Stoughton, C., Allyn Smyth, J., et~al. 2001, ApJ, 553, 184

Chiba, M., \& Beers, T. C., 2000, AJ, 119, 2843

Dalcanton, J.~J., \& Bernstein, R.~A. 2002, AJ, 124, 1328 

Edvardsson, B., Anderson, J., Gustafsson, B., Lambert, D.~L., Nissen, P.~E. \& Tomkin, J., 1993, A\&A 275, 101

Feltzing, S., Bensby, T., \& Lundstr\"{o}m, I. 2003, A\&A, 397, 1

Freeman, K.~C., 1991 in {\it{Dynamics of Disk Galaxies}}, ed. B. Sundelius, (G\"{o}teborgs University and Chalmer University of Technology, G\"{o}teborg Sweden), p.15

Freeman, K.~C. \& Bland-Hawthorn, J. 2002 ARA\&A, 40 487

Fuhrmann, K., 1998, A\&A, 338, 161


Gilmore, G., Wyse, R.~F.~G. \& Norris, J.~E. 2002 ApJ 574, 39

Gilmore, G. 2003, in {\it{Galaxy Evolution: Theory and Observations}} eds V. Avila-Reese, C. Firmani, C. Frenk \& C. Allen Rev Mex AA, CS, vol. 17, 149 

Gilmore, G., Wyse, R. F. G., \& Kuijken, K. 1989, ARA\&A, 27, 555

Gilmore, G., Wyse, R. F. G., \& Jones, J. B. 1995, AJ, 109, 3

Gilmore, G., \& Reid, N. 1983, MNRAS, 202, 1025

Gilmore, G., \& Wyse, R.~F.~G. 1985, AJ, 90, 2015

Gratton, R. G., Carretta, E., Matteucci, F., \& Sneden, C. 2000, A\&A, 358, 671

Habing, H.~J. 1988, A\&A, 200, 40

Quinn, P. J., Hernquist, L., Fullagar, D. P.  1993, ApJ 403, 74

Jones, B.~J.~T., \& Wyse, R.~F.~G.  1983 A\&A, 120, 165

Kawata, D. \& Gibson, B.~K. 2003,  2003, MNRAS, 340, 908

Katz, N. 1992, ApJ, 391, 502 

Katz, N., \& Gunn, J.~E., 1991, ApJ, 377, 565 


Larson, R.~B. 1976, MNRAS 176, 31

Mashonkina, L., \& Gehren, T. 2001, A\&A, 376, 232

Mashonkina, L., Gehren, T., Travaglio, C., \& Borkova, T. 2003, A\&A, 397, 275

Martin, J.~C., Morrison, H.~L. 1998 AJ, 116, 1724

Nordestr\"{o}m et~al. 2004 A\&A, in press

Norris, John, 1987 in {\it{The Galaxy}}, ed. G. Gilmore, B.~Carswell, p. 297, Dordrecht: Reidel


Nissen P. E., Schuster W. J., 1997, A\&A, 326, 751

Ojha, D.~K. 2001, MNRAS, 322, 426  

Phelps, S., Meisenheimer, K., Fuchs, B., Wolf, C. \& Jahreiss, H., 1999, in {\it{Galaxy Evolution: Connecting the Distant Universe with Local Fossil Record}}, Obs de Meudon, 1998.


Prochaska, J. X., Naumov, S. O., Carney, B. W., McWilliam, A., \& Wolfe, A. M. 2000, ApJ, 120, 2513 

Quillen A.~C., and Garnett D. 2001. in {\it{Galaxy Disks and Disk Galaxies}}, ASP Conf. Ser., ed. G Jose, SJ Funes, EM Corsini, 230:87-88. San Francisco: Publ. Astron. Soc. Pac. 


Reddy, B. E., Tomkin, J., Lambert, D. L., \& Allende Prieto, C. 2003, MNRAS, 340, 304 

Robin, A.~C., Haywood, H., Cr\'{e}z\'{e}, M., Ojha, D.~K., Bienaym\'{e}, O. 1996, A\&A, 305, 125

Shetrone, M.~D., Côté, P. \& Sargent, W. L. W. 2001, ApJ 548 592

Shetrone, M., Venn, K.~A., Tolstoy, E., Primas, F., Hill, V. \& Kaufer, A.  2003 AJ 125 684

Schr\"{o}der, K.~P.  \& Pagel, B.~E.~J.   2003, MNRAS, 343, 1231

Spitzer, L. \& Schwarzschild, M.  1953 ApJ, 118, 106

Statler, Thomas S., 1988 ApJ, 331, 71

Str\"{o}mgren, B. 1987 in  {\it{The Galaxy}}, ed. G. Gilmore, B.~Carswell, p. 297, Dordrecht: Reidel

Tautai\u{s}ien\.{e}, G., Edvardsson, B., Tuominen, I. \& Ilyin, I. 2001, A\&A, 380, 578

Thacker, R.~J. \& Couchman, H.~M.~P. 2000, ApJ, 545, 728

Wyse, R.~F.~G. \& Gilmore, G. 1988, Astron J. 95, 1404

Wyse, R.~F.~G. \& Gilmore, G. 1995, AJ 110, 2771

Wyse R.~F.~G., 2000, in {\it{The Galactic Halo: From Globular Clusters to Field Stars}}, ed. A. Noel et al. (Li\'{e}ge: Inst. d'Astrophys. Géophys.), 305 

Wyse R.~F.~G., 2004, in {\it{`The Local Group as an Astrophysical Laboratory'}}, ed.  M. Livio (Cambridge, Cambridge University Press)


\end{references}
\end{document}